\documentclass[aps,prd,showpacs,preprintnumbers,nofootinbib,twocolumn]{revtex4}

\usepackage{bm}
\usepackage{amssymb,psfig}

\newcommand{\be}{\begin{equation}}
\newcommand{\e}{\end{equation}}
\newcommand{\bear}{\begin{eqnarray}}
\newcommand{\ear}{\end{eqnarray}}
\newcommand{\nline}{\nonumber \\}
\newcommand{\f}{\frac}

\newcommand{\sech}{{\rm sech}}

\begin{document}

\title{Concept of temperature in multi-horizon spacetimes: Analysis of Schwarzschild-De Sitter metric}

\author{T. Roy Choudhury}
\affiliation{SISSA/ISAS, via Beirut 2-4, 34014 Trieste, Italy}
\email[]{chou@sissa.it}
\homepage[]{http://www.sissa.it/~chou}
\author{T. Padmanabhan}
\affiliation{IUCAA, Ganeshkhind, Pune 411 007, India}

\date{\today}

\begin{abstract}
In case of spacetimes with single horizon, there exist several well-established
procedures for relating the surface gravity of the horizon to 
a thermodynamic temperature. Such procedures, however, cannot  
be extended in a 
straightforward manner when a spacetime has multiple horizons. In 
particular, it is not clear whether there exists a notion 
of global temperature characterizing the multi-horizon spacetimes. 
We examine the conditions under which    a global temperature 
can exist for a 
spacetime with two horizons using the example of Schwarzschild-De Sitter (SDS)
spacetime.
We systematically extend different procedures (like the expectation value of stress tensor,
response of particle detectors, periodicity in the Euclidean time etc.) for identifying a temperature
in the case of spacetimes with single horizon to the SDS spacetime. This analysis is 
facilitated by using a global coordinate chart which covers the entire SDS manifold.
We find that all the procedures lead to a consistent picture characterized by the following
features: (a) In general, SDS spacetime behaves like a non-equilibrium system 
characterized by two temperatures. (b) It is not possible to associate a global temperature with SDS spacetime except when the ratio of the two surface gravities is rational (c) Even when the ratio of the two surface gravities is rational, the thermal nature depends on the coordinate chart used. There exists
a global coordinate chart in which there is global equilibrium temperature while there exist
other charts in which SDS behaves as though it has two different temperatures. The coordinate dependence of the thermal nature is reminiscent of the flat spacetime in Minkowski and Rindler 
coordinate charts. The implications are discussed.
\end{abstract}
\maketitle

\section{Introduction}

It is possible to associate a notion of 
a thermodynamic temperature 
 with metrics having a single horizon. For example,
        the general class of spacetimes described 
by spherically symmetric metrics
of the form
\be
ds^2 = f(r) dt^2 - [f(r)]^{-1} dr^2 - r^2 [d\theta^2 + \sin^2\theta d\phi^2]
\label{eq:genmet}
\e
with $f(r)$ having a, single, simple zero at $r=r_0$ [i.e.,
$f(r) \simeq f'(r_0) (r - r_0)$ near $r=r_0$], 
have a fairly straightforward thermodynamic interpretation.
In fact, it can be shown that the Einstein's equations
can be expressed in the form of a 
thermodynamic relation $T dS=dE - P dV$ for such spacetimes
\cite{paddy-horizons,padmanabhan03c}, with 
the temperature being determined by the surface gravity of the horizon:
\be
\kappa = \f{1}{2} |f'(r_0)|
\label{eq:kappa}
\e
The most familiar metric amongst the above 
class is, of course, the Schwarzschild metric with a black hole event 
horizon, with the temperature being directly related to the mass of the 
black hole.
The De Sitter metric can also be analysed in a similar manner and one  can 
again identify  a unique  temperature for  the  metric \cite{gh77}. However,  since (i) De Sitter
spacetime is not asymptotically flat and (ii) the De Sitter horizon is observer dependent,
certain new difficulties arise in this case. In particular, concept  like ``evaporation'' of the 
cosmological horizon is  not very obvious unlike  in the case of ordinary black 
holes (Some of these issues are discussed in e.g., \cite{paddy-horizons,padmanabhan03c}). 
But  these  issues have nothing to do with the existence of
{\it multiple} horizons in spacetimes and hence are beyond the scope 
of this paper.

A completely different class of conceptual and mathematical difficulties 
arise while dealing with spacetimes having multiple horizons which we shall analyse in this paper.
The simplest spacetime with multiple horizons is that 
of a black hole in a spacetime with a cosmological 
constant,  
described by the Schwarzschild-De Sitter (SDS) metric
\cite{complex-path,tadaki,mu91,sds,mkni98}.
The metric is characterized by the presence of a black hole event horizon
and a cosmological horizon. 
In recent times, studying such a spacetime has acquired 
further significance because 
of the cosmological observations suggesting the existence of 
a non-zero positive cosmological constant 
(\cite{pag++99}; for reviews, see \cite{lambdareviews}).
While the observations can be explained by a wide class of models 
(see, e.g \cite{pc-sn}), including
those in which the cosmic equation of state can depend on spatial scale (see e.g.,
\cite{pc-tachyon}), 
virtually all these models approach the De Sitter (DS) spacetime at late times 
and at large scales. 
Thus any black hole which forms in the real universe with a cosmological constant
provides an idealization of a SDS spacetime.

The SDS metric has the same form (in one coordinate chart) as the metric in
(\ref{eq:genmet}) but with a $f(r)$ that has two simple zeros at $r=r_\pm$ and two surface
gravities $\kappa_\pm = (1/2) |f'(r_\pm)| $. Naively, one could associate
two different temperatures to the black hole and  cosmological horizon 
using the two different   surface gravities. Since the surface gravities are (generically)
different, the spacetime behaves like a system with two temperatures --- somewhat
like a solid with its edges kept at two thermal baths of two temperatures.
 In that case, there 
will be no well-defined notion of 
global temperature associated with the spacetime \cite{mkni98,medved02}. 
While this sounds plausible, one must bear in mind that the notion of a temperature
in spacetimes with horizon is neither local nor coordinate independent. Hence it is not clear whether one can associate
two separate temperatures with the two horizons.
On the other hand,
there are also some indications that  one can associate  a single, effective 
temperature  with the SDS spacetime  at least in 
some special cases and possibly in specific coordinate charts \cite{tadaki,mkni98,complex-path}.
This suggests that both the viewpoints could be correct but in different coordinate charts.
It should be stressed that the temperature is not a property of the spacetime geometry
in general but depends on the coordinate chart used by a class of observers \cite{paddy-horizons,padmanabhan03c}.
[The best known example is the flat spacetime itself which acquires an observer dependent  horizon
and temperature in the Rindler coordinate chart.]
 In view of these complications, we feel it is worth analysing the temperature of SDS spacetime in some
 detail, which we attempt to do in this paper. 

Our approach will be as follows.  We will simplify mathematical complexities by working with 
1+1 spacetimes since it is well known that the issues we are attempting to address exist even
in two dimensions. We shall then use the standard procedures which lead to the concept of 
a temperature in the case of single horizon metrics to the SDS metric. This analysis shows
that the naive notion of associating two different temperatures with the two horizons is indeed
justifiable at least in some approximate sense. One of the approaches, based on periodicity of 
Euclidean time indicates that there could exist a notion of global temperature in SDS when the 
ratio of surface gravities is a rational number. To investigate this issue carefully, we use
a global coordinate chart which covers the entire SDS manifold and show that \emph{in this
coordinate chart} there does exist a global notion of temperature for the SDS metric when
the ratio of surface gravities is rational.
However, even in this case, the spacetime behaves as though it has two different temperatures
in certain coordinate charts while it has one equilibrium temperature in another global coordinate
chart. This is reminiscent of flat spacetime which exhibits different thermal characteristics 
in different coordinate chart. The implications of this result are discussed in the last section.

\section{General expressions for a metric with 
single horizon}

In this section, we shall briefly review the basic results for a 
spherically symmetric 
spacetime with single horizon, described by the metric (\ref{eq:genmet}). 
In what follows, the angular coordinates $(\theta, \phi)$
do not play any important role, thus allowing us 
to work in the 1+1 dimensional $(t,r)$ subspace. The metric can then be 
written in a conformally flat form by introducing 
the ``tortoise coordinate'' ($r^*$) and the associated null coordinates ($u,v$)
defined by:
\begin{equation}
r^* = \int \frac{d r}{f(r)}; \qquad u = t - r^*, \quad v = t + r^*.
\end{equation}
The metric becomes
the form 
\begin{equation}
d s^2 = f[r(u - v)] \,  du dv.
\end{equation} 
In this form, the metric --- written 
in terms of the $(t,r^*)$, or the $(u,v)$ coordinates --- is 
singular at the horizon where $f(r)$ has a simple zero. To regularize the metric
at the horizon and to remove the singularity, we need to introduce the 
 conventional Kruskal coordinates defined by
\be
U = -\f{1}{\varkappa} {\rm e}^{-\varkappa u},
V = \f{1}{\varkappa} {\rm e}^{\varkappa v}
\e
where 
\be
\varkappa \equiv \f{1}{2} f'(r_0)
\e
(The  
$\varkappa$ can either be positive or negative -- for 
example, it is positive for 
the Schwarzschild metric but is negative for the De Sitter metric. 
On the contrary, the 
surface gravity, which is defined in equation (\ref{eq:kappa}) as
$\kappa = |\varkappa|$, is positive definite.) 
The metric in terms of the Kruskal coordinates is
\begin{equation}
d s^2 = f[r(U - V)] {\rm e}^{-2 \varkappa r^*(U - V)} \ dU dV,
\end{equation} 
which, near the horizon, is free from singularities
as $d s^2 \approx 2 \kappa r_0 dU dV$. 

We shall now briefly summarize the different procedures that can  be
 used    to associate the notion of 
a temperature  with the above metric. 
(Detailed discussion of these and similar procedures for probing the vacuum structure can be found in \cite{bd82,sp02})

\subsection{Expectation value of the energy-momentum tensor}

One can study some of the thermodynamic properties of the spacetime by 
computing the expectation values of the energy-momentum tensor
$T^i_k$ for the corresponding metric. 
In order to calculate the expectation value of 
$T^i_k$ of the matter field in a given spacetime, one needs to 
define a quantum state of the system. Even if one takes it 
to be the vacuum state, there still remains an ambiguity in choosing 
the vacuum in curved spacetime, and thus the expectation 
value of $T^i_k$ will depend on the choice of the vacuum state.

In an 1+1 dimensional conformally flat spacetime (which is the case 
we are interested in), the mode functions 
are simple plane waves, thus simplifying the calculations substantially. 
For a spacetime with single horizon, there can be (at least) three natural
choices of the vacuum state corresponding to different sets of ingoing and 
outgoing modes. The outgoing and incoming modes of the form 
$(4 \pi \omega)^{-1/2} [{\rm e}^{- i \omega u},{\rm e}^{- i \omega v}]$ 
define the Boulware vacuum (B) \cite{boulware75}, while 
$(4 \pi \omega)^{-1/2} [{\rm e}^{- i \omega U},{\rm e}^{- i \omega V}]$
define the Hartle-Hawking vacuum (H) \cite{hh76}. 
The third vacuum is defined as 
$(4 \pi \omega)^{-1/2} [{\rm e}^{- i \omega U},{\rm e}^{- i \omega v}]$ 
and is called the Unruh vacuum (U) \cite{unruh76}.

Usually one is interested in the expectation values of the 
stress tensor as will be 
measured by a freely-falling inertial observer. Since near the horizon, the 
coordinate system $(u,v)$ is singular, the coordinates 
appropriate for the inertial observer will 
be the Kruskal coordinates $(U,V)$. 
Following standard calculations, the expectation values 
of the $T_{UU}, T_{VV}, T_{UV}$ components of the stress tensor 
in the three vacua states are \cite{cf77}
\bear
\langle B | T_{UU} | B \rangle &=& \f{1}{96 \pi \kappa^2 U^2} 
\left[f f'' - \f{1}{2} f'^2\right] \nline
\langle H | T_{UU} | H \rangle &=& \langle U | T_{UU} | U \rangle
=  \langle B | T_{UU} | B \rangle + \f{1}{48 \pi U^2}\nline
\langle B | T_{VV} | B \rangle &=& \langle U | T_{VV} | U \rangle = 
\f{1}{96 \pi \kappa^2 V^2} 
\left[f f'' - \f{1}{2} f'^2\right] \nline
\langle H | T_{VV} | H \rangle &=& \langle B | T_{VV} | B \rangle + 
\f{1}{48 \pi V^2}\nline
\langle B | T_{UV} | B \rangle &=& 
\langle H | T_{UV} | H \rangle =
\langle U | T_{UV} | U \rangle \nline
&=&
-\f{1}{96 \pi \kappa^2 UV} f f''\nline
\ear
It is easy to show that $\langle B | T_{UU} | B \rangle$ diverges (as $U^{-2}$) 
near the horizon, while $\langle H | T_{UU} | H \rangle$ is 
finite there. Furthermore, the 
difference between the Boulware and Hartle-Hawking 
vacua signifies a presence of 
a thermal bath in the latter vacuum. This point is clear 
from the relation
\bear
\langle H | T^t_t | H \rangle - \langle B | T^t_t | B \rangle &=& 
-(\langle H | T^r_r | H \rangle - \langle B | T^r_r | B \rangle)
\nline
& = & \kappa^2/(24 \pi f).
\ear
The temperature of the thermal bath in the Hartle-Hawking vacuum 
can simply be read off from the 
above expression and is seen to be $\kappa/2 \pi$. On the 
other hand, the Unruh vacuum possesses a completely new 
property which is not present in the other two vacua. Note that  
for Unruh vacuum, $\langle U | T_{UU} | U \rangle 
\neq \langle U | T_{UU} | U \rangle$, which indicates the 
presence of a non-zero 
flux of energy. In fact, a straightforward calculation will 
show that
\bear
\langle B | T^r_t | B \rangle &=& \langle H | T^r_t | H \rangle = 0 \nline
\langle U | T^r_t | U \rangle &=& \f{\kappa^2}{48 \pi}
\ear
The form of the energy flux in the Unruh vacuum suggests a blackbody 
emission (in 1+1 dimensions) at a temperature $\kappa/2 \pi$.

The above calculations, although done in only 
1+1 dimensions, show that there is a clear association between 
the surface gravity of the horizon and a 
 temperature defined in the spacetime. A full 3+1 dimensional 
calculation, which is technically much more difficult to tackle, is 
however not expected to alter the above conclusions in general.

\subsection{Response of particle detectors}

The results of the above calculations 
can be compared with 
 the response of a model particle detector in the different 
vacua states. For a massless scalar field in 1+1 dimensions, 
the response function is related to the standard 
Wightman functions $D^+(x,x')$. When the detector is at a fixed 
spatial distance $r$, the response functions per unit 
proper time for the three vacua are given by \cite{bd82}
\bear
{\cal F}_B(E)/\mbox{unit proper time} &= & 0, \nline
{\cal F}_H(E)/\mbox{unit proper time} &\propto & 
\f{1}{E \left({\rm e}^{2 \pi \sqrt{f(r)} E/\kappa} - 1\right)} \nline
{\cal F}_U(E)/\mbox{unit proper time} &\propto& 
\f{1}{ E \left({\rm e}^{2 \pi \sqrt{f(r)} E/\kappa} - 1\right)} \nline
\ear
These expressions, which are essentially of the Planckian form in 2-dimensions,  
indicate that the detector detects a thermal bath at 
apparent temperature $\kappa/2 \pi$ in the Hartle-Hawking vacuum, while 
it detects a flux of particles with the same temperature in the Unruh 
vacuum. This is what we had concluded in the previous section too.

\subsection{Periodicity in the Euclidean time}

Another way of relating the  notion of temperature with the horizon  
is
by considering the periodicity in the Euclidean time coordinate. 
The basic idea is to analytically continue the metric to imaginary values of $t$.
Setting
$
t_E \equiv i t
$
we get
\be
-d s^2 =  \left[f(r) d t_E^2 + \f{d r^2}{f(r)}\right]
\e
The behaviour of this metric near the horizon $r = r_0$
is seen to be of the Rindler form
\be
-d s^2 \approx 
\left[d \rho^2 + \left(\f{\varkappa \beta}{2 \pi}\right)^2 \rho^2 
d \phi^2\right]
\e
where we have defined two new coordinates
$\rho = \int d r/\sqrt{f(r)}$ and $\phi = 2 \pi t_E/\beta$.
In general, this metric has the form of a 2-dimensional 
flat spacetime written in polar coordinates. In order to avoid the conical singularity at the origin, we
need to maintain the  periodicity of the angular coordinate $\phi$;
that is, we  
require that $t_E$ should have a period $n \beta$, where 
\be
\beta = \f{2 \pi}{\kappa}
\e
and $n$ is a (positive) integer. 
The minimum possible period of the Euclidean time $t_E$, 
given by $2 \pi/\kappa$, is 
precisely equal to the inverse 
of the temperature corresponding to the horizon.

The periodicity in $t_E$ can also be seen from the relation between 
the $(t,r)$ and the Kruskal coordinates $(T = [U + V]/2, R = [V - U]/2)$:
\be
T = \f{e^{\varkappa r^*}}{\varkappa} \sinh\varkappa t, 
R = \f{e^{\varkappa r^*}}{\varkappa} \cosh\varkappa t
\e
The relations between the Euclidean time coordinates in these 
two systems are
\be
T_E = \f{e^{\varkappa r^*}}{\kappa} \sin\kappa t_E
\label{ttrel}
\e
which shows that $T_E$ is periodic function of $t_E$ with the (minimum) 
period of 
$t_E$ being given by $2 \pi n/\kappa$. Similar conclusions can 
be drawn by considering the spatial coordinate $R$ too. 
This periodicity will be shown 
by any analytic function of the coordinates $(T_E, R)$ over the whole 
manifold.
In particular, the Greens function defined over the 
entire spacetime will be an 
analytic function of $(T_E, R)$ and hence will be periodic 
in the imaginary time coordinate. One can then 
analytically continue this 
Euclidean Greens function and 
obtain the Feynman 
propagator in the original $(t,r)$ space which also be periodic in the 
imaginary time. In general, 
this periodicity of the propagator can be shown to be 
a characteristic of a thermal state (see, for example, \cite{wald84}) with 
a temperature given by the inverse of the period $\beta^{-1}$. 
Thus the period of the Euclidean time $t_E$ is seen to be directly 
related to the temperature of the spacetime.

We shall now apply the above procedures for associating a temperature with the horizon
to the case of 
SDS metric. 
 
\section{Schwarzschild-De Sitter (S-DS) spacetime}

We now extend the formalisms of the previous section to the 
Schwarzschild-De Sitter (S-DS) spacetime, described by the metric
\be
d s^2 = \left(1 - \f{2 M}{r} - H^2 r^2\right) d t^2 - 
\left(1 - \f{2 M}{r} - H^2 r^2\right)^{-1} d r^2 
\label{eq:sdsmet}
\e
where we have omitted the angular coordinates. 
This metric has 
two horizons at $r_-$ and $r_+$ -- they are 
the black hole event horizon and the cosmological horizon respectively.
Let us denote the surface gravities associated with the 
two horizons by $\kappa_-$ and $\kappa_+$ respectively. 
(The detailed expressions for these quantities are 
given in Appendix A.)
The metric can be written in the conformal form by introducing the 
usual `tortoise coordinate' $r^* $
and the null coordinates  $u = t - r^*, v = t + r^*$. Until this point, the analysis follows 
exactly as in the case of the single horizon.

In the case of spacetimes with single horizon, one next introduces the Kruskal
coordinates thereby obtaining a non-singular coordinate chart which covers
the whole manifold.  Note, however, that this transformation necessarily involves
the surface gravity which is different from for the two horizons.
 When the spacetime has more than 
one horizons, one has to use a Kruskal coordinate patch for 
each of them -- no single patch of usual Kruskal-type coordinates
can cover both the 
horizons. For example, to remove the singularity near the 
black-hole horizon $r_-$, introduce the coordinates
\be
U_- = -\f{1}{\kappa_-} {\rm e}^{-\kappa_- u}; ~~ 
V_- = \f{1}{\kappa_-} {\rm e}^{\kappa_- v}
\e
so that the metric becomes
\be
d s^2 = \f{2M}{r} 
\left(1 - \f{r}{r_+}\right)^{1 + \f{\kappa_-}{\kappa_+}}
\left(1 + \f{r}{r_- + r_+}\right)
^{2 - \f{\kappa_-}{\kappa_+}} d U_- d V_-
\label{eq:ds2du-dv-}
\e
Although this is non-singular near the black hole horizon $r_-$, it is 
clear that the 
singularity at the other horizon $r_+$ still prevails. Thus the 
coordinates $(U_-, V_-)$ are not defined 
for the region $r > r_+$. Near the cosmological horizon $r_+$, 
we can introduce another set of Kruskal 
coordinates 
\be
U_+ = \f{1}{\kappa_+} {\rm e}^{\kappa_+ u}; ~~ 
V_+ = -\f{1}{\kappa_+} {\rm e}^{-\kappa_+ v}
\e
so that the metric becomes
\be
d s^2 = \f{2M}{r} 
\left(\f{r}{r_-} - 1\right)^{1 + \f{\kappa_+}{\kappa_-}}
\left(1 + \f{r}{r_- + r_+}\right)
^{2 - \f{\kappa_+}{\kappa_-}} d U_+ d V_+
\label{eq:ds2du+dv+}
\e
This metric, in turn,  is singular at $r = r_-$ and hence these coordinates 
cannot be extended to the region $r < r_-$.
In the region of overlap ($r_- < r < r_+$), where both the 
coordinate patches are well defined, they are related to each other by
\be
(-\kappa_- U_-)^{-\f{1}{\kappa_-}} = (\kappa_+ U_+)^{\f{1}{\kappa_+}}, ~
(\kappa_- V_-)^{-\f{1}{\kappa_-}} = (-\kappa_+ V_+)^{\f{1}{\kappa_+}}
\e

These considerations already show that one will expect non-trivial differences
between the case of single horizon and multi-horizon scenarios.
We shall now extend the concepts introduced for a single horizon to 
SDS metric and see whether there is any global temperature associated 
with the spacetime.

\subsection{Expectation values of the stress tensor}

The calculations performed for the expectation values of the stress tensor 
in the single horizon case can be 
extended to the SDS spacetime by identifying the corresponding vacuum states. 
There is no ambiguity in the Boulware vacuum, defined in terms of the 
modes $(4 \pi \omega)^{-1/2} [{\rm e}^{- i \omega u},{\rm e}^{- i \omega v}]$. 
The calculations and the expressions are 
are identical to the single horizon case. Since the 
$(u,v)$ coordinate system is badly behaved at both the horizons, 
the expectation values for an inertial observer 
will diverge at both the horizons.

The usual 
Hartle-Hawking vacuum is defined in terms of the Kruskal coordinates. 
Since we have two different patches of Kruskal coordinates, it 
is possible to define two separate Hartle-Hawking vacua. Let us call them 
as
\begin{eqnarray}
 &&{\rm H}_-: 
(4 \pi \omega)^{-1/2} [{\rm e}^{- i \omega U_-},{\rm e}^{- i \omega V_-}] \nonumber\\
&&{\rm H}_+: 
(4 \pi \omega)^{-1/2} [{\rm e}^{- i \omega U_+},{\rm e}^{- i \omega V_+}].
\end{eqnarray} 
For these vacua, 
one can trivially 
extend the calculations for the single horizon case. 
As expected,
for an inertial observer in the H$_-$ vacuum, the expectation values 
will be finite at the black hole horizon $r_-$, but will 
diverge at the cosmological horizon $r_+$. Similar conclusions will 
hold for H$_+$ too. It also 
turns out that, as expected, the H$_-$ vacuum corresponds to a thermal bath 
at a temperature $\kappa_-/2 \pi$, while the H$_+$ vacuum 
gives a thermal bath 
at a temperature $\kappa_+/2 \pi$.

Similarly, there will now be two different sets of Unruh vacua too. 
Let us call them as 
\begin{eqnarray}
&&{\rm U}_-:
(4 \pi \omega)^{-1/2} [{\rm e}^{- i \omega U_-},{\rm e}^{- i \omega v}]\nonumber \\
&&{\rm U}_+:
(4 \pi \omega)^{-1/2} [{\rm e}^{- i \omega u},{\rm e}^{- i \omega V_+}]. 
\end{eqnarray}
The only non-trivial result one obtains from the standard Unruh 
vacuum is the presence of a thermal flux. 
Following straightforward calculations, one can show in this case
\be
\langle U_- | T^r_t | U_- \rangle = \f{\kappa_-^2}{48 \pi},
\langle U_+ | T^r_t | U_+ \rangle = -\f{\kappa_+^2}{48 \pi}
\e
This indicates that each vacuum represents 
a flux with a temperature associated with the corresponding horizon. 

All the results mentioned above are direct generalizations of the corresponding 
results in the case single horizon. There is, however, one non-trivial
new result that arises in the SDS spacetime. We can  define another
 non-trivial vacuum state  for 
a spacetime with two horizons which has no analogue for the single 
horizon spacetimes. It is very much like the 
standard Unruh vacuum, and is defined as
\begin{equation}
{\rm U}_{-+}:
(4 \pi \omega)^{-1/2} [{\rm e}^{- i \omega U_-},{\rm e}^{- i \omega V_+}].
\end{equation} 
This has the same outgoing modes as the standard U$_-$ vacuum, while 
the ingoing modes are like the U$_+$.
The flux obtained from this vacuum state has the expression
\be
\langle U_{-+} | T_{xt} | U_{-+} \rangle = 
\f{\varkappa_-^2 - \varkappa_+^2}{48 \pi} 
\e
This indicates the presence of two oppositely driven thermal 
fluxes with different temperatures -- one flowing 
from the black hole horizon with temperature $\kappa_-/2 \pi$, while other 
flowing from the cosmological horizon with temperature $\kappa_+/2 \pi$. 
Just as the standard Unruh vacuum is useful for studying collapse of matter 
in Schwarzschild spacetime \cite{unruh76}, it has been argued that 
the U$_{-+}$ is appropriate for describing collapse 
of matter in a spacetime with cosmological horizon \cite{mu91,tadaki}.

It turns out from the above analysis 
that there is no notion of a global temperature which is 
characteristic of the entire spacetime. We shall now see that the 
same conclusion can be drawn by studying the response of particle 
detectors in different vacua.

\subsection{Detector response}

We have already defined five vacuum states for the 
SDS metric in the previous section. The calculation for the 
detector response functions in this spacetime is quite
straightforward. We just mention the final results:
\bear
\f{{\cal F}_B(E)}{\mbox{unit proper time}} &=& 0, \nline
\f{{\cal F}_{H_-}(E)}{\mbox{unit proper time}} &\propto& 
\f{1}{E \left({\rm e}^{2 \pi \sqrt{f(r)} E/\kappa_-} - 1\right)} \nline
\f{{\cal F}_{H_+}(E)}{\mbox{unit proper time}} &\propto& 
\f{1}{E \left({\rm e}^{2 \pi \sqrt{f(r)} E/\kappa_+} - 1\right)} \nline
\f{{\cal F}_{U_-}(E)}{\mbox{unit proper time}} &\propto& 
\f{1}{ E \left({\rm e}^{2 \pi \sqrt{f(r)} E/\kappa_-} - 1\right)} \nline
\f{{\cal F}_{U_+}(E)}{\mbox{unit proper time}} &\propto& 
\f{1}{E \left({\rm e}^{2 \pi \sqrt{f(r)} E/\kappa_+} - 1\right)} \nline
\label{eq:sdsdetresp}
\ear
The main conclusions drawn in the previous section go through unchanged in each of these cases.
The Boulware vacuum shows no flux, while the Hartle-Hawking and Unruh vacua exhibit
two different temperatures depending on the state that is chosen.

The new feature arises in the  $U_{-+}$ state. Here we find that:
\bear
\f{{\cal F}_{U_{-+}}(E)}{\mbox{unit proper time}} &\propto& 
\f{1}{ E} \left[\f{1}{({\rm e}^{2 \pi \sqrt{f(r)} E/\kappa_-} - 1)}
\right].
\nline
&+& \left.\f{1}{({\rm e}^{2 \pi \sqrt{f(r)} E/\kappa_+} - 1)} \right]
\ear
That is,
the detector in the 
U$_{-+}$ vacuum register {\it simultaneously} two different thermal fluxes of temperatures 
$\kappa_-/2 \pi$ and $\kappa_+/2 \pi$.  Since the superposition of two Planckian spectra cannot be mapped to a  Planck spectrum of single temperature, there is no 
indication of a global temperature associated with the spacetime. 
[At the Rayleigh limit the Planck spectrum is proportional to the temperature and, in this limit, the effective temperature is simply the sum of the two temperatures. At the other extreme of high frequencies, the higher
temperature will dominate. So clearly, the notion of effective temperature is frequency dependent.]

\subsection{Euclidean time and periodicity}

All the above results were tuned to the static coordinate system [$r^*,t$] and the {\it two}
Kruskal coordinate systems obtained from them. Since all these coordinate patches are singular, all of them are conceptually comparable to Rindler coordinate system in flat spacetime and we need to find the analogue of global Minkowski coordinate system. In this aspect, the SDS differs drastically from either Schwarzschild or De Sitter, in either of which, the Kruskal coordinates cover the global manifold and 
are non-singular.
One might, therefore,  argue that since we have not used a global coordinate system 
which can cover both the horizons, it is understandable that we do not have a
 global temperature. We shall now take up 
this issue.

The first indication, that a global notion of temperature might exist, arises when we study the periodicity in
the Euclidean time.
As in the case of the single horizon, one can write the 
metric in the Rindler form near each of the horizons. Near 
$r = r_{\pm}$, the metric becomes
\be
d s^2 \approx -\left[d \rho^2 + 
\left(\f{\kappa_{\pm} \beta}{2 \pi}\right)^2 \rho^2 
d \phi^2\right]
\e
As we have seen before, 
in order to maintain the periodicity of the angular coordinate 
$\phi$ near $r = r_-$, we must have  
$\beta = 2 \pi n_-/\kappa_-$, where 
$n_-$ is a positive integer. Similar argument near $r_+$ shows that 
one has to have $\beta = 2 \pi n_+/\kappa_+$, where $n_+$ is some 
other integer. Thus, if one wants to have a globally defined 
period for $t_E$, then the following condition must hold true
\be
\f{\kappa_+}{\kappa_-} = \f{n_+}{n_-} 
\e
i.e., the ratio of the two surface gravities should be rational. 
(From now on, we assume that $n_-$ and $n_+$ are relatively prime 
integers.)
We thus seem to arrive at a global notion of a thermal temperature 
provided we impose a condition on the ratio of the 
surface gravities. 
In 
case the ratio is not rational, there would not be any globally defined 
period of $t_E$ and the Euclidean metric will have a conical singularity
at (at least) one of the horizons.

\section{ Analysis in a global coordinate chart}

The above conclusion above can verified by 
considering the periodicity of the Greens function which is analytic 
over the whole spacetime. However, for this 
purpose, one first requires a global coordinate patch 
which can cover the whole manifold. 
To settle the above issue in a direct manner, we shall analyse the problem in a global coordinate system which covers the entire manifold.

In the case of single horizon, 
we have seen that the Kruskal coordinates are adequate for this purpose. 
On the contrary, for a spacetime with multiple horizons, one has to 
look for something else. It turns out that there is a {\it family} 
of coordinate systems which are globally well defined, covers the full manifold and are non-singular. We take a 
particular one of them in this paper. The details of the 
global coordinates $(\bar{T}, \bar{R})$ --- which are quite complicated algebraically ---
are discussed in Appendix B. 
The only relation we need here is the dependences of the 
global Euclidean time coordinate $\bar{T}_E = i \bar{T}$ and $\bar{R}$ 
on the original $t_E$ [which is the analogue of equation (\ref{ttrel})]. The relation for $\bar{T}_E$ is given by
\bear
&\bar{T_E} = \f{1}{\kappa_-} 
\f{\tan \left({\rm e}^{\kappa_- r^*} \sin \kappa_- t_E \right)  
\sech^2 \left({\rm e}^{\kappa_- r^*} \cos \kappa_- t_E \right)}
{1 + \tanh^2 \left({\rm e}^{\kappa_- r^*} \cos \kappa_- t_E  \right) 
\tan^2 \left({\rm e}^{\kappa_- r^*} \sin \kappa_- t_E \right)}&
\nline
& + \f{1}{\kappa_+} 
\f{\tan \left({\rm e}^{-\kappa_+ r^*} \sin \kappa_+ t_E \right) 
\sech^2 \left({\rm e}^{-\kappa_+ r^*} \cos \kappa_+ t_E \right)}
{1 + \tanh^2 \left({\rm e}^{-\kappa_+ r^*} \cos \kappa_+ t_E \right) 
\tan^2 \left({\rm e}^{-\kappa_+ r^*} \sin \kappa_+ t_E \right)}&
\ear
Note that the first term periodic when $t_E$ has 
a period $\beta = 2 \pi n_-/\kappa_-$, while the 
second term is periodic when $\beta = 2 \pi n_+/\kappa_+$. This implies 
that $\bar{T}_E$ is periodic only when 
$\kappa_+/\kappa_- = n_+/n_-$. (The same conclusion 
can be drawn by considering the expression for 
$\bar{R}$.) This, in turn, implies that the 
Euclidean Greens function is periodic in $t_E$ only when 
the  ratio $\kappa_+/\kappa_-$ is rational. In that case, 
the Feynman propagator describes a thermal state having a temperature 
\begin{equation}
\beta^{-1} = \kappa_-/(2 \pi n_-) = \kappa_+/(2 \pi n_+).
\end{equation}  
In case 
the ratio $\kappa_+/\kappa_-$ is not rational, the
Euclidean Greens function $\bar{T}_E$ is not a periodic function of 
$t_E$ and the propagator does not characterize a 
thermal state any more and hence there 
is no notion of a globally defined temperature.

When we have $\kappa_+/\kappa_- = n_+/n_-$, and that there is 
a global temperature $\kappa_-/(2 \pi n_-) = \kappa_+/(2 \pi n_+)$, one 
could also verify that  the usual limits exist 
when $M \to 0$ or $H \to 0$. When the black hole mass
vanishes ($M \to 0$), we have $\kappa_- \to \infty$ and the period of 
the term containing $\kappa_-$ will go to zero. This essentially means that 
the relevant term will oscillate rapidly, and in the limit of its frequency 
going to infinity, it will complete infinite number of periods by the time 
the other term completes one period. In that case, the period 
of $\bar{T}_E$, and hence the temperature of the spacetime,  
will be determined by the period of the term containing $\kappa_+$, and 
the temperature will be simply be the horizon temperature $\kappa_+/2\pi$. 
In the other limit when $H \to 0$, 
we have $\kappa_+ \to 0$. 
As $\kappa_+$ vanishes, the term containing $\kappa_+$ will stop 
oscillating and will essentially 
be a constant. In this case the period will be determined by $\kappa_-$ and 
we will get back the black hole temperature.
The difference in the manner in which the limits are obtained has nothing to do with our analysis of the problem. It merely reflects the following well-known, but curious, feature: When $H\to 0$ in the pure De Sitter metric,
the spacetime becomes flat and the surface gravity [and temperature] vanishes.  But when $M\to 0$ in the Schwarzschild metric, the spacetime becomes flat {\it but the surface gravity and the temperature diverge}.

Even when $\kappa_+/\kappa_- = n_+/n_-$, the notion of a global temperature is coordinate dependent.
This is most easily seen from the analysis in Section III. If we choose the $H_\pm$ or $U_\pm,U_{-+}$
vacuum states, the stress tensor expectation values or the detector response will still lead to the results
in equations (\ref{eq:sdsdetresp}). 
There are two temperatures and fluxes corresponding to two temperatures, except that one temperature is  rational multiple of the other. On the other hand, quantum field theory in the global coordinate chart
will describe a system with an effective temperature $\beta^{-1}=\kappa_-/(2 \pi n_-) = \kappa_+/(2 \pi n_+)$. [This can be explicitly shown using the coordinate chart developed in Appendix  B but is, of course, obvious from the Euclidean periodicity arguments]. Thus we see that both the claims in the literature (``the SDS has two temperatures'' and ``SDS has an effective single temperature'') are correct but applies to different coordinate charts. This is similar, in a limited sense, to the flat spacetime appearing as having a global zero temperature in the Minkowski chart but exhibiting a non-zero temperature in the Rindler chart.

\section{Conclusions}

Our analysis shows that SDS spacetime possesses a dual thermal interpretation in a manner similar to flat spacetime in Minkowski and Rindler coordinate charts, if the surface gravities satisfy the condition 
$\kappa_+/\kappa_- = n_+/n_-$. In this case, there is a global coordinate chart in which the metric can be described as having a temperature $\beta^{-1}=\kappa_-/(2 \pi n_-) = \kappa_+/(2 \pi n_+)$. More conventional coordinate charts lead to the interpretation of SDS having two different temperatures. We stress that while this result is quite interesting (and new, as far as the authors are aware of), such coordinate dependence is well
known in quantum field theory in curved spacetime. The only difference between the SDS situation and 
the Minkowski/Rindler situation is that in the latter the global temperature is zero while in SDS the global temperature is non-zero. 

The rational numbers are infinitely dense in the space of 
real numbers and in practical sense the condition on the 
surface gravities may {\it not} impose any condition on $M$ and $H$ at all -- 
however the very fact that demanding equilibrium in a spacetime 
imposes a restriction on the surface gravities is an interesting 
point just as a matter of principle.
It would be interesting to examine the possibility of whether any 
semiclassical calculations can lead to such quantization of surface gravities. 
It has been suggested from two different classes of semiclassical gravity 
calculations that the areas of the horizons might be 
quantized [in units of (Planck length)$^2$] 
(for reviews, see \cite{rovelli98,padmanabhan03c}). 
In that case we have
$r_+^2/r_-^2 = N_+/N_-$, where $N_{\pm}$ are (relatively prime) 
positive integers. 
The condition on the ratio of surface gravities is then
\be
\f{\kappa_+}{\kappa_-} = \f{N_- + 2 \sqrt{N_+ N_-}}{N_+ + 2 \sqrt{N_+ N_-}}
\e
The ratio $\kappa_+/\kappa_-$ can still be an irrational number, hence
the quantization of areas does {\it not} necessarily imply the 
existence of thermal equilibrium. It is clear from the 
above expression that $\kappa_+/\kappa_-$ will be rational 
only when $\sqrt{N_+/N_-}$ is rational, i.e., 
$N_+/N_- = N_1^2/N_2^2$, where $N_1$ and $N_2$ form another set 
of relatively prime integers. Hence, for the existence of a 
global temperature one requires that 
$r_+/r_- = N_1/N_2$. This means that 
the quantization of areas is not sufficient, there 
must be further restrictions on the horizon radii.
{\it In particular, it is adequate if the radii of horizons are quantized in the units of Planck length
which, of course, is consistent with the notion of area quantization.}
In such case, 
the condition on $MH$ will be
\be
M H = \f{N_2}{2} \f{N_1 N_2 (N_1 + N_2)}
{(N_1^2 + N_1 N_2 + N_2^2)^{3/2}}
\label{cond}
\e
The implications of this result are under investigation.

The situation is more unclear when the ratio of surface gravities is not a rational number. The global coordinate system, defined in Appendix B, still exists covering the whole manifold but the metric in this coordinate system does not lead to any thermal interpretation. The other, singular coordinate charts, of course, lead to the conventional view of two different temperatures for the SDS. The somewhat disturbing feature in this case is that, the Euclidean metric, obtained by the analytic continuation in $t$ will necessarily have a conical singularity.
Hence, a non-singular Euclidean quantum field theory does not exist in this case. It is, however, unclear whether one should {\it demand} the existence of the non-singular Euclidean field theory while working on a given curved spacetime. After all, an arbitrary, time dependent background spacetime may not even have a 
Euclidean continuation, let alone a non-singular one.  But if we make such a demand then we obtain certain bizarre conclusions. For example, if the universe has a cosmological constant, then any black hole that forms in it  must have a mass which satisfies the quantization condition in equation (\ref{cond}).

Finally, we mention that the thermal behaviour of horizons is closely related to the quasi-normal modes
(QNM)
as pointed out in recent analyses of QNM's using Born approximation \cite{pc-qnm}. This investigation shows that the QNM's of the SDS spacetime arises essentially from those of the Schwarzschild metric. In the study
performed in this paper, however, both the horizons contribute in equal footing. It will be, therefore, interesting to analyse the semiclassical wave modes in the global coordinate system and compare them with 
the results in the singular coordinate charts. This, and related issues, are under study.

\appendix

\section{Horizons, surface gravities and their relations in the SDS spacetime}

In this appendix, let us discuss the basic properties of the 
SDS metric.
The SDS spacetime, described by 
(\ref{eq:sdsmet}),  has two horizons at $r_-$ and $r_+$ (with $r_- < r_+$):
\be
r_- = \f{2}{\sqrt{3} H} \sin\left(\f{\theta}{3}\right), ~
r_+ = \f{2}{\sqrt{3} H} \sin\left(\f{\theta + 2 \pi}{3}\right)
\label{eq:x+-}
\e
where
\be
\sin \theta = 3^{3/2} M H; ~~ 0 < \theta < \f{\pi}{2}
\label{eq:theta}
\e
and we have assumed $0 < M H < 3^{-3/2}$. As discussed, 
the horizon at 
$r_-$ is called the black hole horizon, while that at 
$r_+$ is called the cosmological horizon.
As $M H \to 0$, one 
obtains the two known limits $r_- \to 2 M$ and $r_+ \to H^{-1}$.
The surface gravity for the two horizons are given by
\bear
\kappa_- &=& \f{H^2}{2 r_-} 
(r_+ - r_-) (r_+ + 2 r_-) \nline
\kappa_+ &=& \f{H^2}{2 r_+} 
(r_+ - r_-) (r_- + 2 r_+)
\label{eq:kappa+-}
\ear
As $M H \to 0$, one 
gets the usual limits $\kappa_- \to (4 M)^{-1}$ and $\kappa_+ \to H$.

The metric can be written in the conformal form by introducing the 
`tortoise coordinate'
\bear
r^* &=& \int d r \left(1 - \f{2 M}{r} - H^2 r^2\right)^{-1} \nline
&=& \f{1}{2 \kappa_-} \ln\left|\f{r}{r_-} - 1\right|
- \f{1}{2 \kappa_+} \ln\left|1 - \f{r}{r_+}\right| \nline
&-& \f{1}{2} \left(\f{1}{\kappa_-} -  \f{1}{\kappa_+} \right) 
\ln\left|\f{r}{r_- + r_+} + 1\right|
\ear
The null coordinates can be defined as $u = t - r^*, v = t + r^*$.

One can, in principle, write the parameters $M$ and $H$ in terms of 
the surface gravities $\kappa_-$ and $\kappa_+$ of the two horizons. 
However, it turns out that this cannot be written in a closed form. 
On the other hand, it is possible to write the combination $M H$ in 
terms of the ratio $\kappa_+/\kappa_-$. One can get from equation
(\ref{eq:kappa+-})
\be
k = \f{1 + 2 x}{x (x + 2)}
\e
where we have defined
\be
x \equiv \f{r_+}{r_-}
\e
and
\be
k \equiv \f{\kappa_+}{\kappa_-}
\e
This can be inverted to obtain
\be
x = \f{1 - k + \sqrt{1 - k + k^2}}{k}
\e
Note that $k = 1$ only if $x = 1$, i.e., the surface gravities
of the two horizons are equal only when the two horizons coincide.
Also, since $x \ge 1$, one gets $k \le 1$. This implies that 
the temperature corresponding to the cosmological horizon 
is always less than that of the black hole horizon.

Now, use equation (\ref{eq:x+-}) to write
\be
x = \f{\sin[(\theta + 2 \pi)/3]}{\sin(\theta/3)}
\e
The terms containing $\theta$ can be directly related to the 
combination $M H$ using equation (\ref{eq:theta}), so that
\be
M H = \f{1}{2} \f{x (1+x)}{(x^2 + x + 1)^{3/2}}
\e
and finally
\be
M H = \f{k}{2} \f{2 - 2 k + k^2 + (2 - k) \sqrt{1 - k + k^2}}
{[2 (1 - k + k^2) + (2 - k) \sqrt{1 - k + k^2}]^{3/2}}
\e

\section{Global coordinates for the SDS spacetime}

\begin{figure*}
\begin{center}
\psfig{figure=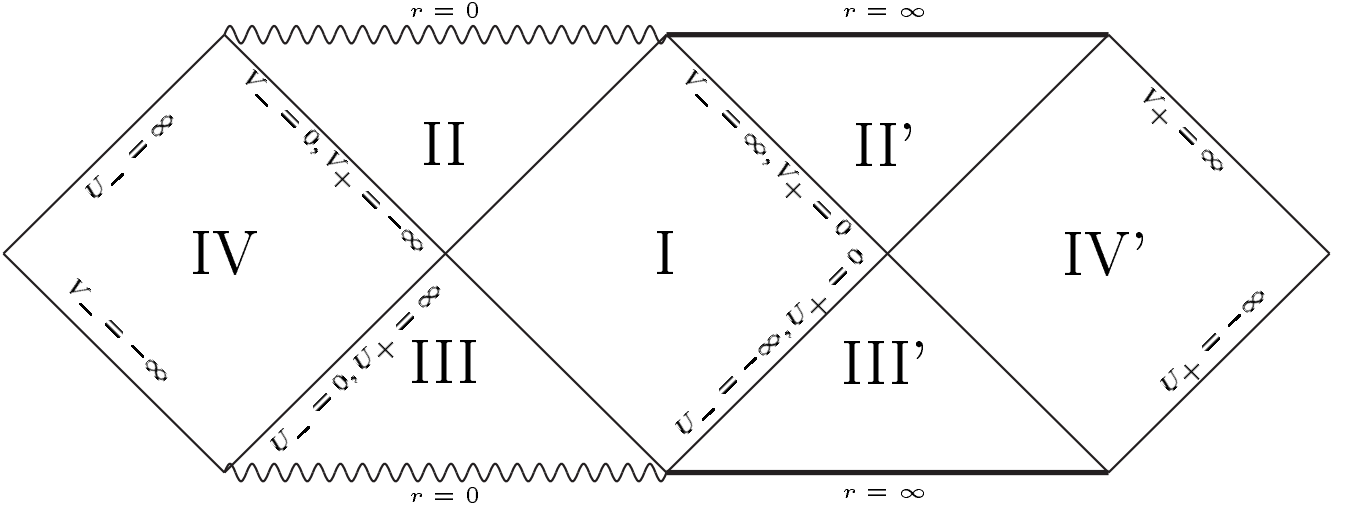,width=\textwidth}
\end{center}
\caption[]
{Penrose diagram for the Schwarzschild-De Sitter spacetime
}
\label{sds-penrose}
\end{figure*}

This appendix discusses the existence and properties of a global
coordinate system in the SDS spacetime. 
As discussed in the main text, 
unlike the case of a metric with single horizon, one cannot
cover the entire SDS manifold with the conventional Kruskal-like 
patches. However, it turns out that there exists a class 
of coordinate systems which are analytic over the whole space.
Let us study one explicit example for this.
The global coordinates $(\bar{U}, \bar{V})$ can be defined in 
each region of the Penrose diagram (see Figure \ref{sds-penrose}) 
in terms of the Kruskal coordinates defined in the corresponding region.
For completeness, we give the detailed expressions for the 
global coordinates in each region:

Region I: $U_- < 0, V_- > 0, U_+ > 0, V_+ < 0$
\bear
\bar{U} &=& \f{1}{\kappa_-} \tanh \kappa_- U_- 
+ \f{1}{\kappa_+} \tanh \kappa_+ U_+ \nline
&=&  \f{1}{\kappa_-} \tanh \kappa_- U_- 
+ \f{1}{\kappa_+} \tanh (-\kappa_- U_-)^{-\f{\kappa_+}{\kappa_-}}
\nline
&=& -\f{1}{\kappa_-} \tanh (\kappa_+ U_+)^{-\f{\kappa_-}{\kappa_+}}
+ \f{1}{\kappa_+} \tanh \kappa_+ U_+ 
\nline
\bar{V} &=& \f{1}{\kappa_-} \tanh \kappa_- V_- 
+ \f{1}{\kappa_+} \tanh \kappa_+ V_+ \nline
&=&  \f{1}{\kappa_-} \tanh \kappa_- V_- 
- \f{1}{\kappa_+} \tanh (\kappa_- V_-)^{-\f{\kappa_+}{\kappa_-}}
\nline
&=& \f{1}{\kappa_-} \tanh (-\kappa_+ V_+)^{-\f{\kappa_-}{\kappa_+}}
+ \f{1}{\kappa_+} \tanh \kappa_+ V_+ 
\ear

Region II: $U_- > 0, V_- > 0$
\bear
\bar{U} &=& \f{1}{\kappa_-} \tanh \kappa_- U_- 
- \f{1}{\kappa_+} \tanh (\kappa_- U_-)^{-\f{\kappa_+}{\kappa_-}}
+\f{2}{\kappa_+}
\nline
\bar{V} &=& \f{1}{\kappa_-} \tanh \kappa_- V_- 
- \f{1}{\kappa_+} \tanh (\kappa_- V_-)^{-\f{\kappa_+}{\kappa_-}}
\nline &&
\ear

Region III: $U_- < 0, V_- < 0$
\bear
\bar{U} &=& \f{1}{\kappa_-} \tanh \kappa_- U_- 
+ \f{1}{\kappa_+} \tanh (-\kappa_- U_-)^{-\f{\kappa_+}{\kappa_-}}
\nline
\bar{V} &=& \f{1}{\kappa_-} \tanh \kappa_- V_- 
+ \f{1}{\kappa_+} \tanh (-\kappa_- V_-)^{-\f{\kappa_+}{\kappa_-}} 
- \f{2}{\kappa_+}
\nline &&
\ear

Region II': $U_+ > 0, V_+ > 0$
\bear
\bar{U} &=& -\f{1}{\kappa_-} \tanh (\kappa_+ U_+)^{-\f{\kappa_-}{\kappa_+}}
+ \f{1}{\kappa_+} \tanh \kappa_+ U_+ 
\nline
\bar{V} &=& -\f{1}{\kappa_-} \tanh (\kappa_+ V_+)^{-\f{\kappa_-}{\kappa_+}}
+ \f{1}{\kappa_+} \tanh \kappa_+ V_+ + \f{2}{\kappa_-}
\nline &&
\ear

Region III':  $U_+ < 0, V_+ < 0$
\bear
\bar{U} &=& \f{1}{\kappa_-} \tanh (-\kappa_+ U_+)^{-\f{\kappa_-}{\kappa_+}}
+ \f{1}{\kappa_+} \tanh \kappa_+ U_+ - \f{2}{\kappa_-}
\nline
\bar{V} &=& \f{1}{\kappa_-} \tanh (-\kappa_+ V_+)^{-\f{\kappa_-}{\kappa_+}}
+ \f{1}{\kappa_+} \tanh \kappa_+ V_+ 
\nline &&
\ear

Region IV: $U_- > 0, V_- < 0$
\bear
\bar{U} &=& \f{1}{\kappa_-} \tanh \kappa_- U_- 
- \f{1}{\kappa_+} \tanh (\kappa_- U_-)^{-\f{\kappa_+}{\kappa_-}}
+\f{2}{\kappa_+}
\nline
\bar{V} &=& \f{1}{\kappa_-} \tanh \kappa_- V_- 
+ \f{1}{\kappa_+} \tanh (-\kappa_- V_-)^{-\f{\kappa_+}{\kappa_-}} 
- \f{2}{\kappa_+}
\nline &&
\ear

Region IV': $U_+ < 0, V_+ > 0$
\bear
\bar{U} &=& \f{1}{\kappa_-} \tanh (-\kappa_+ U_+)^{-\f{\kappa_-}{\kappa_+}}
+ \f{1}{\kappa_+} \tanh \kappa_+ U_+ - \f{2}{\kappa_-}
\nline
\bar{V} &=& -\f{1}{\kappa_-} \tanh (\kappa_+ V_+)^{-\f{\kappa_-}{\kappa_+}}
+ \f{1}{\kappa_+} \tanh \kappa_+ V_+ + \f{2}{\kappa_-}
\nline &&
\ear

One can see that the relations are quite similar in various regions. 
It is clear from the expressions that the global coordinates 
$(\bar{U}, \bar{V})$ reduce to 
the Kruskal patch $(U_-, V_-)$ near the black hole horizon, and to 
$(U_+, V_+)$ near the cosmological horizon. In fact, 
any coordinate
patch with such a property will be regular at both the horizons and can act 
as a global coordinate system.

We shall now discuss the various properties of this coordinate 
system. To avoid unnecessary complications, 
from now on, we shall concentrate on two or three particular regions, 
namely, I and II (and II' if required). The relations for all the 
other regions can be trivially extended. Note that the boundary between 
I and II denotes the future black hole horizon ($r = r_-, U_- = 0$) while 
that between 
I and II' denotes the future cosmological horizon ($r = r_+, V_+ = 0$).

\subsection{Continuity}

It is possible to show that $\bar{U}$ and $\bar{V}$ are continuous 
over the whole spacetime by checking their values at the boundaries. 
For example, consider the boundary between I and II, where $U_- = 0$. 
The expressions for $\bar{V}$ are identical in these two regions, and 
hence the continuity of $\bar{V}$ is obvious. 
The value of $\bar{U}$ near the boundary in region I is 
\be
\lim_{U_- \to 0^-} \bar{U} = \f{1}{\kappa_+} + U_- - \f{\kappa_-^2 U_-^3}{3}
\e
while that in region II is
\be
\lim_{U_- \to 0^+} \bar{U} = \f{1}{\kappa_+} + U_- - \f{\kappa_-^2 U_-^3}{3}
\e
This shows that 
$\bar{U}$ and its derivative are continuous across the boundary. 
Similarly, near the boundary between I and II', we have $V_+ = 0$. 
This time 
the expressions for $\bar{U}$ are identical in these two regions, and 
hence the continuity of $\bar{U}$ is obvious. 
The value of $\bar{V}$ near the boundary in region I is 
\be
\lim_{V_+ \to 0^+} \bar{V} = \f{1}{\kappa_-} + V_+ - \f{\kappa_+^2 V_+^3}{3}
\e
while that in region II' is
\be
\lim_{V_+ \to 0^-} \bar{V} = \f{1}{\kappa_-} + V_+ - \f{\kappa_+^2 V_+^3}{3}
\e
and the continuity follows.
Similar proof can be given for all the other cases.

\subsection{Explicit form of the metric}

The metric can be written in  terms of the global coordinates as
\be
d s^2 = C(\bar{U},\bar{V}) d\bar{U} d\bar{V}
\e
The quantity $C(\bar{U},\bar{V})$ can be written explicitly in various regions.
For example, in region I, use equation (\ref{eq:ds2du-dv-}) to write 
it in terms of $r$ and $(U_-, V_-)$
\bear
C(\bar{U},\bar{V}) &=& \f{2M}{r} 
\left(1 - \f{r}{r_+}\right)^{1 + \f{\kappa_-}{\kappa_+}}
\left(1 + \f{r}{r_- + r_+}\right)
^{2 - \f{\kappa_-}{\kappa_+}}
\nline
&\times& 
\left(\f{d \bar{U}}{d U_-} \f{d \bar{V}}{d V_-}\right)^{-1} 
\ear
which gives
\bear
&&\!\!\!\!\!\! C(\bar{U},\bar{V}) = \f{2M}{r} 
\left(1 - \f{r}{r_+}\right)^{1 + \f{\kappa_-}{\kappa_+}}
\left(1 + \f{r}{r_- + r_+}\right)
^{2 - \f{\kappa_-}{\kappa_+}} \nline
&&\!\!\!\!\!\! \times
\left[\sech^2(\kappa_- U_-) + (-\kappa_- U_-)^{-\f{\kappa_+}{\kappa_-} - 1}
\sech^2(-\kappa_- U_-)^{-\f{\kappa_+}{\kappa_-}} \right]^{-1} \nline
&&\!\!\!\!\!\! \times
\left[\sech^2(\kappa_- V_-) + (\kappa_- V_-)^{-\f{\kappa_+}{\kappa_-} - 1}
\sech^2(\kappa_- V_-)^{-\f{\kappa_+}{\kappa_-}} \right]^{-1}
\label{eq:cuv_I}
\ear
[Similarly, one can start with equation (\ref{eq:ds2du+dv+}) to write 
$C(\bar{U},\bar{V})$  in terms of $r$ and $(U_+, V_+)$.]
Note that $C(\bar{U},\bar{V})$ 
is positive definite in the region, as required. It might seem 
from the way it is written that it vanishes at the cosmological 
horizon $r = r_+$. However, a straightforward 
calculation shows that the conformal factor near $r = r_+$ is
\be
C(\bar{U},\bar{V}) \approx \f{2M}{r} 
\left(1 + \f{r}{r_- + r_+}\right)
^{2 - \f{\kappa_+}{\kappa_-}} 
\left(\f{r}{r_-} - 1\right)^{\f{\kappa_+}{\kappa_-} + 1} 
\e
which is non-vanishing and non-singular. For completeness, let us 
determine its behaviour near the other horizon $r = r_-$
\be
C(\bar{U},\bar{V}) \approx \f{2M}{r} 
\left(1 - \f{r}{r_+}\right)^{1 + \f{\kappa_-}{\kappa_+}}
\left(1 + \f{r}{r_- + r_+}\right)
^{2 - \f{\kappa_-}{\kappa_+}} 
\e

Let us also write the explicit form of $C(\bar{U},\bar{V})$ in region II. 
Note that only the coordinates $(U_-, V_-)$ are defined in this region. 
Thus one can use equation (\ref{eq:ds2du-dv-}) to get
\bear
&&\!\!\!\!\!\! C(\bar{U},\bar{V}) = \f{2M}{r} 
\left(1 - \f{r}{r_+}\right)^{1 + \f{\kappa_-}{\kappa_+}}
\left(1 + \f{r}{r_- + r_+}\right)
^{2 - \f{\kappa_-}{\kappa_+}} \nline
&&\!\!\!\!\!\! \times
\left[\sech^2(\kappa_- U_-) + (\kappa_- U_-)^{-\f{\kappa_+}{\kappa_-} - 1}
\sech^2(\kappa_- U_-)^{-\f{\kappa_+}{\kappa_-}} \right]^{-1} \nline
&&\!\!\!\!\!\! \times
\left[\sech^2(\kappa_- V_-) + (\kappa_- V_-)^{-\f{\kappa_+}{\kappa_-} - 1}
\sech^2(\kappa_- V_-)^{-\f{\kappa_+}{\kappa_-}} \right]^{-1}
\nline&&
\ear
As required, $C(\bar{U},\bar{V})$ 
is positive definite in the region too. Near the 
horizon $r = r_-$, one can perform a calculation similar to the 
case in region I, and show that
\be
C(\bar{U},\bar{V}) \approx \f{2M}{r} 
\left(1 - \f{r}{r_+}\right)^{1 + \f{\kappa_-}{\kappa_+}}
\left(1 + \f{r}{r_- + r_+}\right)
^{2 - \f{\kappa_-}{\kappa_+}} 
\e
which is identical to the corresponding expression in region I. 
Thus the metric is continuous and analytic across the horizon. 

One can perform identical calculations for the other regions and 
find that the global coordinates $(\bar{U},\bar{V})$ actually 
cover the whole space, and the metric is free from singularities at 
both the horizons.

\subsection{The limiting cases}

At this point, let us check the limiting case of $M H \to 0$ for the 
global coordinates. 
In region I, the global coordinates, written in terms of $(t,r)$ coordinates 
are given by
\bear
\bar{U} &=& -\f{1}{\kappa_-} \tanh 
\left[{\rm e}^{-\kappa_- (t - r^*)} 
\right]
+ \f{1}{\kappa_+} \tanh 
\left[{\rm e}^{\kappa_+ (t - r^*)}\right]
\nline
&=&-\f{1}{\kappa_-} \tanh 
\left[{\rm e}^{-\kappa_- t}
\sqrt{\f{r}{r_-} - 1} 
\left(1 - \f{r}{r_+}\right)^{-\f{\kappa_-}{2 \kappa_+}} \right.
\nline
& &\times
\left.
\left(1 + \f{r}{r_- + r_+}\right)^{(\f{\kappa_-}{\kappa_+} - 1)/2} \right]
\nline
&+& \f{1}{\kappa_+} \tanh 
\left[{\rm e}^{\kappa_+ t}
\left(\f{r}{r_-} - 1\right)^{-\f{\kappa_+}{2 \kappa_-}}
\sqrt{1 - \f{r}{r_+}} \right.
\nline
& &\times
\left.
\left(1 + \f{r}{r_- + r_+}\right)^{(\f{\kappa_+}{\kappa_-} - 1)/2}\right]
\ear
and
\bear
\bar{V} &=& \f{1}{\kappa_-} \tanh 
\left[{\rm e}^{\kappa_- (t + r^*)} 
\right]
- \f{1}{\kappa_+} \tanh 
\left[{\rm e}^{-\kappa_+ (t + r^*)}\right]
\nline
&=& \f{1}{\kappa_-} \tanh 
\left[{\rm e}^{\kappa_- t}
\sqrt{\f{r}{r_-} - 1}
\left(1 - \f{r}{r_+}\right)^{-\f{\kappa_-}{2 \kappa_+}} \right.
\nline
& &\times
\left.
\left(1 + \f{r}{r_- + r_+}\right)^{(\f{\kappa_-}{\kappa_+} - 1)/2} \right]
\nline
&-& \f{1}{\kappa_+} \tanh 
\left[{\rm e}^{-\kappa_+ t}
\left(\f{r}{r_-} - 1\right)^{-\f{\kappa_+}{2 \kappa_-}}
\sqrt{1 - \f{r}{r_+}} \right.
\nline
& &\times
\left.
\left(1 + \f{r}{r_- + r_+}\right)^{(\f{\kappa_+}{\kappa_-} - 1)/2}\right]
\ear
As $M H \to 0$, we have 
$r_- \to 2 M, r_+ \to H^{-1}, \kappa_- \to (4 M)^{-1}$ and $\kappa_+ \to H$.
Then, in the limit, the expressions become
\bear
\bar{U} 
&\approx&
-4 M \tanh 
\left[{\rm e}^{-t/4 M}
\sqrt{\f{r}{2 M} - 1} ~
\left(\f{1 + H r}{1 - H r}\right)^{1/8 M H} \right]
\nline
&+& \f{1}{H} \tanh 
\left[{\rm e}^{H t}
\sqrt{\f{1 - H r}{1 + H r}} ~
\right]
\ear
Now, as $H \to 0$, we have
\be
\bar{U} \approx -4 M \tanh
\left[{\rm e}^{-t/4 M} {\rm e}^{r/4M}
\sqrt{\f{r}{2 M} - 1} \right]
\e
where we have used $(1 + H r)^{(1/8 MH)} \approx \exp(r/8 M)$ and
$(1 - H r)^{(-1/8 MH)} \approx \exp(r/8 M)$ as $H \to 0$. Thus the global 
coordinate reduces to tanh of the usual Kruskal coordinates in the 
Schwarzschild spacetime. Similarly, when $M \to 0$, we have
\be
\bar{U} \approx \f{1}{H} \tanh 
\left[{\rm e}^{H t}
\sqrt{\f{1 - H r}{1 + H r}} ~
\right]
\e
which is the tanh of the usual Kruskal coordinates in the 
De Sitter spacetime. Similar calculations can be done for $\bar{V}$ in region 
I, and then one can extend the calculations to all the other regions.

\subsection{Imaginary time and periodicity}

Let us introduce the coordinates $(\bar{T}, \bar{R})$:
\be
\bar{T} = \f{\bar{U} + \bar{V}}{2},~~
\bar{R} = \f{\bar{V} - \bar{U}}{2}
\e
Now, analytically continue 
the $\bar{T}$ coordinate to imaginary space and define 
the Euclidean time coordinates 
\be 
\bar{T}_E = i \bar{T}
\e
One would like to express the set $(\bar{R}, \bar{T}_E)$ in terms 
of $(r,t_E)$. Note that in region I, putting $t = -i t_E$ in the expressions 
for $\bar{U}$ and $\bar{V}$ shows that 
$\bar{U} = -\bar{V}^{\dagger}$. Then
$\bar{T}_E = i (\bar{U} + \bar{V})/2 = i (\bar{V} - \bar{V}^{\dagger})/2
= -\Im(\bar{V})$ and 
similarly $\bar{R} = \Re(\bar{V})$.
A straightforward calculation shows that 
\bear
&\bar{T_E} = \f{1}{\kappa_-} 
\f{\tan \left({\rm e}^{\kappa_- r^*} \sin \kappa_- t_E \right)  
\sech^2 \left({\rm e}^{\kappa_- r^*} \cos \kappa_- t_E \right)}
{1 + \tanh^2 \left({\rm e}^{\kappa_- r^*} \cos \kappa_- t_E  \right) 
\tan^2 \left({\rm e}^{\kappa_- r^*} \sin \kappa_- t_E \right)}&
\nline
& + \f{1}{\kappa_+} 
\f{\tan \left({\rm e}^{-\kappa_+ r^*} \sin \kappa_+ t_E \right) 
\sech^2 \left({\rm e}^{-\kappa_+ r^*} \cos \kappa_+ t_E \right)}
{1 + \tanh^2 \left({\rm e}^{-\kappa_+ r^*} \cos \kappa_+ t_E \right) 
\tan^2 \left({\rm e}^{-\kappa_+ r^*} \sin \kappa_+ t_E \right)}&
\nline
&\bar{R} = \f{1}{\kappa_-} 
\f{\sec^2 \left({\rm e}^{\kappa_- r^*} \sin \kappa_- t_E \right)  
\tanh \left({\rm e}^{\kappa_- r^*} \cos \kappa_- t_E \right)}
{1 + \tanh^2 \left({\rm e}^{\kappa_- r^*} \cos \kappa_- t_E  \right) 
\tan^2 \left({\rm e}^{\kappa_- r^*} \sin \kappa_- t_E \right)}&
\nline
& - \f{1}{\kappa_+} 
\f{\sec^2 \left({\rm e}^{-\kappa_+ r^*} \sin \kappa_+ t_E \right) 
\tanh^2 \left({\rm e}^{-\kappa_+ r^*} \cos \kappa_+ t_E \right)}
{1 + \tanh^2 \left({\rm e}^{-\kappa_+ r^*} \cos \kappa_+ t_E \right) 
\tan^2 \left({\rm e}^{-\kappa_+ r^*} \sin \kappa_+ t_E \right)}&
\ear
This shows that $\bar{T_E}$ and $\bar{R}$ is periodic only when 
$\kappa_+/\kappa_- = n_+/n_-$, and the period of $t_E$ 
is given by $\beta = 2 \pi n_+/\kappa_+ = 2 \pi n_-/\kappa_-$.

One can show that the conformal factor 
$C(\bar{U},\bar{V})$ is positive definite in region I 
when written in terms of the imaginary time  
(this follows from equation (\ref{eq:cuv_I}) and from 
the fact that $U_- = -V_-^{\dagger}$ when 
written in terms of $t_E$).

\bibliography{astropap}

\begin{thebibliography}{59}
\expandafter\ifx\csname natexlab\endcsname\relax\def\natexlab#1{#1}\fi
\expandafter\ifx\csname bibnamefont\endcsname\relax
  \def\bibnamefont#1{#1}\fi
\expandafter\ifx\csname bibfnamefont\endcsname\relax
  \def\bibfnamefont#1{#1}\fi
\expandafter\ifx\csname citenamefont\endcsname\relax
  \def\citenamefont#1{#1}\fi
\expandafter\ifx\csname url\endcsname\relax
  \def\url#1{\texttt{#1}}\fi
\expandafter\ifx\csname urlprefix\endcsname\relax\def\urlprefix{URL }\fi
\providecommand{\bibinfo}[2]{#2}
\providecommand{\eprint}[2][]{\url{#2}}

\bibitem[{\citenamefont{Padmanabhan}(2002{\natexlab{a}})}]{paddy-horizons}
\bibinfo{author}{\bibfnamefont{T.}~\bibnamefont{Padmanabhan}},
  \bibinfo{journal}{Class. Quant. Grav.} \textbf{\bibinfo{volume}{19}},
  \bibinfo{pages}{5387} (\bibinfo{year}{2002}{\natexlab{a}});
\bibinfo{author}{\bibfnamefont{T.}~\bibnamefont{Padmanabhan}},
  \bibinfo{journal}{Mod. Phys. Lett.} \textbf{\bibinfo{volume}{A17}},
  \bibinfo{pages}{923} (\bibinfo{year}{2002}{\natexlab{b}}),
  \eprint{gr-qc/0202078};
\bibinfo{author}{\bibfnamefont{T.}~\bibnamefont{Padmanabhan}},
  \bibinfo{journal}{Class. Quant. Grav.} \textbf{\bibinfo{volume}{19}},
  \bibinfo{pages}{3551} (\bibinfo{year}{2002}{\natexlab{c}}),
  \eprint{gr-qc/0110046}.

\bibitem[{\citenamefont{Padmanabhan}(2003{\natexlab{a}})}]{padmanabhan03c}
\bibinfo{author}{\bibfnamefont{T.}~\bibnamefont{Padmanabhan}},
  \bibinfo{howpublished}{To appear in Rev. Mod. Phys., gr-qc/0311036}
  (\bibinfo{year}{2003}).

\bibitem[{\citenamefont{Gibbons and Hawking}(1977)}]{gh77}
\bibinfo{author}{\bibfnamefont{G.~W.} \bibnamefont{Gibbons}} \bibnamefont{and}
  \bibinfo{author}{\bibfnamefont{S.~W.} \bibnamefont{Hawking}},
  \bibinfo{journal}{Phys. Rev.} \textbf{\bibinfo{volume}{D15}},
  \bibinfo{pages}{2738} (\bibinfo{year}{1977}).

\bibitem[{\citenamefont{Shankaranarayanan}(2003)}]{complex-path}
\bibinfo{author}{\bibfnamefont{S.}~\bibnamefont{Shankaranarayanan}},
  \bibinfo{journal}{Phys. Rev.} \textbf{\bibinfo{volume}{D67}},
  \bibinfo{pages}{084026} (\bibinfo{year}{2003}), \eprint{gr-qc/0301090};
\bibinfo{author}{\bibfnamefont{K.}~\bibnamefont{Srinivasan}} \bibnamefont{and}
  \bibinfo{author}{\bibfnamefont{T.}~\bibnamefont{Padmanabhan}},
  \bibinfo{journal}{Phys. Rev.} \textbf{\bibinfo{volume}{D60}},
  \bibinfo{pages}{024007} (\bibinfo{year}{1999}), \eprint{gr-qc/9812028};
\bibinfo{author}{\bibfnamefont{S.}~\bibnamefont{Shankaranarayanan}},
  \bibinfo{author}{\bibfnamefont{K.}~\bibnamefont{Srinivasan}},
  \bibnamefont{and}
  \bibinfo{author}{\bibfnamefont{T.}~\bibnamefont{Padmanabhan}},
  \bibinfo{journal}{Mod. Phys. Lett.} \textbf{\bibinfo{volume}{A16}},
  \bibinfo{pages}{571} (\bibinfo{year}{2001}), \eprint{gr-qc/0007022}.

\bibitem[{\citenamefont{Tadaki and Takagi}(1990{\natexlab{a}})}]{tadaki}
\bibinfo{author}{\bibfnamefont{S.-I.} \bibnamefont{Tadaki}} \bibnamefont{and}
  \bibinfo{author}{\bibfnamefont{S.}~\bibnamefont{Takagi}},
  \bibinfo{journal}{Prog. Theor. Phys.} \textbf{\bibinfo{volume}{83}},
  \bibinfo{pages}{941} (\bibinfo{year}{1990}{\natexlab{a}});
\bibinfo{author}{\bibfnamefont{S.}~\bibnamefont{Tadaki}} \bibnamefont{and}
  \bibinfo{author}{\bibfnamefont{S.}~\bibnamefont{Takagi}},
  \bibinfo{journal}{Prog. Theor. Phys.} \textbf{\bibinfo{volume}{83}},
  \bibinfo{pages}{1126} (\bibinfo{year}{1990}{\natexlab{b}}).

\bibitem[{\citenamefont{Markovic and Unruh}(1991)}]{mu91}
\bibinfo{author}{\bibfnamefont{D.}~\bibnamefont{Markovic}} \bibnamefont{and}
  \bibinfo{author}{\bibfnamefont{W.~G.} \bibnamefont{Unruh}},
  \bibinfo{journal}{Phys. Rev.} \textbf{\bibinfo{volume}{D43}},
  \bibinfo{pages}{332} (\bibinfo{year}{1991}).

\bibitem[{\citenamefont{Bousso and Hawking}(1998)}]{sds}
\bibinfo{author}{\bibfnamefont{R.}~\bibnamefont{Bousso}} \bibnamefont{and}
  \bibinfo{author}{\bibfnamefont{S.~W.} \bibnamefont{Hawking}},
  \bibinfo{journal}{Phys. Rev.} \textbf{\bibinfo{volume}{D57}},
  \bibinfo{pages}{2436} (\bibinfo{year}{1998}), \eprint{hep-th/9709224};
\bibinfo{author}{\bibfnamefont{F.-L.} \bibnamefont{Lin}} \bibnamefont{and}
  \bibinfo{author}{\bibfnamefont{C.}~\bibnamefont{Soo}},
  \bibinfo{journal}{Class. Quant. Grav.} \textbf{\bibinfo{volume}{16}},
  \bibinfo{pages}{551} (\bibinfo{year}{1999}), \eprint{gr-qc/9708049};
\bibinfo{author}{\bibfnamefont{S.}~\bibnamefont{Nojiri}} \bibnamefont{and}
  \bibinfo{author}{\bibfnamefont{S.~D.} \bibnamefont{Odintsov}},
  \bibinfo{journal}{Phys. Rev.} \textbf{\bibinfo{volume}{D59}},
  \bibinfo{pages}{044026} (\bibinfo{year}{1999}), \eprint{hep-th/9804033};
\bibinfo{author}{\bibfnamefont{S.}~\bibnamefont{Nojiri}} \bibnamefont{and}
  \bibinfo{author}{\bibfnamefont{S.~D.} \bibnamefont{Odintsov}},
  \bibinfo{journal}{Int. J. Mod. Phys.} \textbf{\bibinfo{volume}{A15}},
  \bibinfo{pages}{989} (\bibinfo{year}{2000}), \eprint{hep-th/9905089};
\bibinfo{author}{\bibfnamefont{Z.~C.} \bibnamefont{Wu}}, \bibinfo{journal}{Gen.
  Rel. Grav.} \textbf{\bibinfo{volume}{32}}, \bibinfo{pages}{1823}
  (\bibinfo{year}{2000}), \eprint{gr-qc/9911078};
\bibinfo{author}{\bibfnamefont{Y.-Q.} \bibnamefont{Wu}},
  \bibinfo{author}{\bibfnamefont{L.-C.} \bibnamefont{Zhang}}, \bibnamefont{and}
  \bibinfo{author}{\bibfnamefont{R.}~\bibnamefont{Zhao}},
  \bibinfo{journal}{Int. J. Theor. Phys.} \textbf{\bibinfo{volume}{40}},
  \bibinfo{pages}{1001} (\bibinfo{year}{2001});
\bibinfo{author}{\bibfnamefont{R.}~\bibnamefont{Zhao}},
  \bibinfo{author}{\bibfnamefont{J.-F.} \bibnamefont{Zhang}}, \bibnamefont{and}
  \bibinfo{author}{\bibfnamefont{L.-C.} \bibnamefont{Zhang}},
  \bibinfo{journal}{Mod. Phys. Lett.} \textbf{\bibinfo{volume}{A16}},
  \bibinfo{pages}{719} (\bibinfo{year}{2001});
\bibinfo{author}{\bibfnamefont{W.~A.} \bibnamefont{Hiscock}},
  \bibinfo{journal}{Phys. Rev.} \textbf{\bibinfo{volume}{D39}},
  \bibinfo{pages}{1067} (\bibinfo{year}{1989});
\bibinfo{author}{\bibfnamefont{S.}~\bibnamefont{Deser}} \bibnamefont{and}
  \bibinfo{author}{\bibfnamefont{O.}~\bibnamefont{Levin}},
  \bibinfo{journal}{Class. Quant. Grav.} \textbf{\bibinfo{volume}{14}},
  \bibinfo{pages}{L163} (\bibinfo{year}{1997}), \eprint{gr-qc/9706018};
\bibinfo{author}{\bibfnamefont{R.}~\bibnamefont{Zhao}},
  \bibinfo{author}{\bibfnamefont{L.-C.} \bibnamefont{Zhang}}, \bibnamefont{and}
  \bibinfo{author}{\bibfnamefont{Z.-G.} \bibnamefont{Li}},
  \bibinfo{journal}{Nuovo Cim.} \textbf{\bibinfo{volume}{B113}},
  \bibinfo{pages}{291} (\bibinfo{year}{1998});
\bibinfo{author}{\bibfnamefont{S.}~\bibnamefont{Deser}} \bibnamefont{and}
  \bibinfo{author}{\bibfnamefont{O.}~\bibnamefont{Levin}},
  \bibinfo{journal}{Phys. Rev.} \textbf{\bibinfo{volume}{D59}},
  \bibinfo{pages}{064004} (\bibinfo{year}{1999}), \eprint{hep-th/9809159};
\bibinfo{author}{\bibfnamefont{Y.~S.} \bibnamefont{Myung}},
  \bibinfo{journal}{Mod. Phys. Lett.} \textbf{\bibinfo{volume}{A16}},
  \bibinfo{pages}{2353} (\bibinfo{year}{2001}), \eprint{hep-th/0110123};
\bibinfo{author}{\bibfnamefont{S.~Q.} \bibnamefont{Wu}} \bibnamefont{and}
  \bibinfo{author}{\bibfnamefont{X.}~\bibnamefont{Cai}},
  \bibinfo{journal}{Nuovo Cim.} \textbf{\bibinfo{volume}{116B}},
  \bibinfo{pages}{907} (\bibinfo{year}{2001}), \eprint{hep-th/0108033};
\bibinfo{author}{\bibfnamefont{R.}~\bibnamefont{Garattini}},
  \bibinfo{journal}{Class. Quant. Grav.} \textbf{\bibinfo{volume}{18}},
  \bibinfo{pages}{571} (\bibinfo{year}{2001}), \eprint{gr-qc/0012078};
\bibinfo{author}{\bibfnamefont{A.~M.} \bibnamefont{Ghezelbash}}
  \bibnamefont{and} \bibinfo{author}{\bibfnamefont{R.~B.} \bibnamefont{Mann}},
  \bibinfo{journal}{JHEP} \textbf{\bibinfo{volume}{01}}, \bibinfo{pages}{005}
  (\bibinfo{year}{2002}), \eprint{hep-th/0111217};
\bibinfo{author}{\bibfnamefont{M.}~\bibnamefont{Cvetic}},
  \bibinfo{author}{\bibfnamefont{S.}~\bibnamefont{Nojiri}}, \bibnamefont{and}
  \bibinfo{author}{\bibfnamefont{S.~D.} \bibnamefont{Odintsov}},
  \bibinfo{journal}{Nucl. Phys.} \textbf{\bibinfo{volume}{B628}},
  \bibinfo{pages}{295} (\bibinfo{year}{2002}), \eprint{hep-th/0112045};
\bibinfo{author}{\bibfnamefont{S.~Q.} \bibnamefont{Wu}} \bibnamefont{and}
  \bibinfo{author}{\bibfnamefont{X.}~\bibnamefont{Cai}}, \bibinfo{journal}{Int.
  J. Theor. Phys.} \textbf{\bibinfo{volume}{41}}, \bibinfo{pages}{559}
  (\bibinfo{year}{2002}), \eprint{gr-qc/0111045};
\bibinfo{author}{\bibfnamefont{U.~H.} \bibnamefont{Danielsson}},
  \bibinfo{journal}{JHEP} \textbf{\bibinfo{volume}{03}}, \bibinfo{pages}{020}
  (\bibinfo{year}{2002}), \eprint{hep-th/0110265};
\bibinfo{author}{\bibfnamefont{S.}~\bibnamefont{Nojiri}},
  \bibinfo{author}{\bibfnamefont{S.~D.} \bibnamefont{Odintsov}},
  \bibnamefont{and} \bibinfo{author}{\bibfnamefont{S.}~\bibnamefont{Ogushi}},
  \bibinfo{journal}{Int. J. Mod. Phys.} \textbf{\bibinfo{volume}{A18}},
  \bibinfo{pages}{3395} (\bibinfo{year}{2003}), \eprint{hep-th/0212047};
\bibinfo{author}{\bibfnamefont{D.}~\bibnamefont{Guido}} \bibnamefont{and}
  \bibinfo{author}{\bibfnamefont{R.}~\bibnamefont{Longo}},
  \bibinfo{journal}{Annales Henri Poincare} \textbf{\bibinfo{volume}{4}},
  \bibinfo{pages}{1169} (\bibinfo{year}{2003}), \eprint{gr-qc/0212025};
\bibinfo{author}{\bibfnamefont{A.}~\bibnamefont{{Gomberoff}}} \bibnamefont{and}
  \bibinfo{author}{\bibfnamefont{C.}~\bibnamefont{{Teitelboim}}},
  \bibinfo{journal}{Phys. Rev. D} \textbf{\bibinfo{volume}{67}},
  \bibinfo{pages}{104024} (\bibinfo{year}{2003});
\bibinfo{author}{\bibfnamefont{A.}~\bibnamefont{Corichi}} \bibnamefont{and}
  \bibinfo{author}{\bibfnamefont{A.}~\bibnamefont{Gomberoff}},
  \bibinfo{journal}{Phys. Rev.} \textbf{\bibinfo{volume}{D69}},
  \bibinfo{pages}{064016} (\bibinfo{year}{2004}), \eprint{hep-th/0311030};
\bibinfo{author}{\bibfnamefont{P.~C.~W.} \bibnamefont{Davies}}
  \bibnamefont{and} \bibinfo{author}{\bibfnamefont{T.~M.} \bibnamefont{Davis}},
  \bibinfo{howpublished}{Preprint: astro-ph/0310522} (\bibinfo{year}{2003});
\bibinfo{author}{\bibfnamefont{T.~M.} \bibnamefont{Davis}},
  \bibinfo{author}{\bibfnamefont{P.~C.~W.} \bibnamefont{Davies}},
  \bibnamefont{and} \bibinfo{author}{\bibfnamefont{C.~H.}
  \bibnamefont{Lineweaver}}, \bibinfo{journal}{Class. Quant. Grav.}
  \textbf{\bibinfo{volume}{20}}, \bibinfo{pages}{2753} (\bibinfo{year}{2003}),
  \eprint{astro-ph/0305121};
\bibinfo{author}{\bibfnamefont{C.}~\bibnamefont{Teitelboim}}
  (\bibinfo{year}{2002}), \eprint{hep-th/0203258};
\bibinfo{author}{\bibfnamefont{Y.-b.} \bibnamefont{Kim}},
  \bibinfo{author}{\bibfnamefont{C.~Y.} \bibnamefont{Oh}}, \bibnamefont{and}
  \bibinfo{author}{\bibfnamefont{N.}~\bibnamefont{Park}}
  (\bibinfo{year}{2002}), \eprint{hep-th/0212326}.

\bibitem[{\citenamefont{{Maeda} et~al.}(1998)\citenamefont{{Maeda}, {Koike},
  {Narita}, and {Ishibashi}}}]{mkni98}
\bibinfo{author}{\bibfnamefont{K.}~\bibnamefont{Maeda}},
  \bibinfo{author}{\bibfnamefont{T.}~\bibnamefont{Koike}},
  \bibinfo{author}{\bibfnamefont{M.}~\bibnamefont{Narita}}, \bibnamefont{and}
  \bibinfo{author}{\bibfnamefont{A.}~\bibnamefont{Ishibashi}},
  \bibinfo{journal}{Phys. Rev.} \textbf{\bibinfo{volume}{D57}},
  \bibinfo{pages}{3503} (\bibinfo{year}{1998}), \eprint{gr-qc/9712029}.

\bibitem[{\citenamefont{{Perlmutter} et~al.}(1999)\citenamefont{{Perlmutter},
  {Aldering}, {Goldhaber}, {Knop}, {Nugent}, {Castro}, {Deustua}, {Fabbro},
  {Goobar}, {Groom} et~al.}}]{pag++99}
\bibinfo{author}{\bibfnamefont{S.}~\bibnamefont{{Perlmutter}}},
  \bibinfo{author}{\bibfnamefont{G.}~\bibnamefont{{Aldering}}},
  \bibinfo{author}{\bibfnamefont{G.}~\bibnamefont{{Goldhaber}}},
  \bibinfo{author}{\bibfnamefont{R.~A.} \bibnamefont{{Knop}}},
  \bibinfo{author}{\bibfnamefont{P.}~\bibnamefont{{Nugent}}},
  \bibinfo{author}{\bibfnamefont{P.~G.} \bibnamefont{{Castro}}},
  \bibinfo{author}{\bibfnamefont{S.}~\bibnamefont{{Deustua}}},
  \bibinfo{author}{\bibfnamefont{S.}~\bibnamefont{{Fabbro}}},
  \bibinfo{author}{\bibfnamefont{A.}~\bibnamefont{{Goobar}}},
  \bibinfo{author}{\bibfnamefont{D.~E.} \bibnamefont{{Groom}}},
  \bibnamefont{et~al.}, \bibinfo{journal}{Astrophys. J.}
  \textbf{\bibinfo{volume}{517}}, \bibinfo{pages}{565} (\bibinfo{year}{1999}).

\bibitem[{\citenamefont{{Padmanabhan}}(2003)}]{lambdareviews}
\bibinfo{author}{\bibfnamefont{T.}~\bibnamefont{{Padmanabhan}}},
  \bibinfo{journal}{Phys. Rept.} \textbf{\bibinfo{volume}{380}},
  \bibinfo{pages}{235} (\bibinfo{year}{2003});
\bibinfo{author}{\bibfnamefont{V.}~\bibnamefont{Sahni}} \bibnamefont{and}
  \bibinfo{author}{\bibfnamefont{A.~A.} \bibnamefont{Starobinsky}},
  \bibinfo{journal}{Int. J. Mod. Phys.} \textbf{\bibinfo{volume}{D9}},
  \bibinfo{pages}{373} (\bibinfo{year}{2000}),
  \eprint[http://arXiv.org/abs]{astro-ph/9904398};
\bibinfo{author}{\bibfnamefont{P.~J.} \bibnamefont{{Peebles}}}
  \bibnamefont{and} \bibinfo{author}{\bibfnamefont{B.}~\bibnamefont{{Ratra}}},
  \bibinfo{journal}{Rev. Mod. Phys.} \textbf{\bibinfo{volume}{75}},
  \bibinfo{pages}{559} (\bibinfo{year}{2003}).

\bibitem[{\citenamefont{{Padmanabhan} and {Choudhury}}(2003)}]{pc-sn}
\bibinfo{author}{\bibfnamefont{T.}~\bibnamefont{{Padmanabhan}}}
  \bibnamefont{and} \bibinfo{author}{\bibfnamefont{T.~R.}
  \bibnamefont{{Choudhury}}}, \bibinfo{journal}{Mon. Not. R. Astron. Soc.}
  \textbf{\bibinfo{volume}{344}}, \bibinfo{pages}{823} (\bibinfo{year}{2003});
\bibinfo{author}{\bibfnamefont{T.~R.} \bibnamefont{Choudhury}}
  \bibnamefont{and}
  \bibinfo{author}{\bibfnamefont{T.}~\bibnamefont{Padmanabhan}},
  \bibinfo{howpublished}{Preprint: astro-ph/0311622} (\bibinfo{year}{2003}).

\bibitem[{\citenamefont{{Padmanabhan} and {Choudhury}}(2002)}]{pc-tachyon}
\bibinfo{author}{\bibfnamefont{T.}~\bibnamefont{{Padmanabhan}}}
  \bibnamefont{and} \bibinfo{author}{\bibfnamefont{T.~R.}
  \bibnamefont{{Choudhury}}}, \bibinfo{journal}{Phys. Rev. D}
  \textbf{\bibinfo{volume}{66}}, \bibinfo{pages}{081301}
  (\bibinfo{year}{2002}), \eprint{hep-th/0205055};
\bibinfo{author}{\bibfnamefont{J.~S.} \bibnamefont{{Bagla}}},
  \bibinfo{author}{\bibfnamefont{H.~K.} \bibnamefont{{Jassal}}},
  \bibnamefont{and}
  \bibinfo{author}{\bibfnamefont{T.}~\bibnamefont{{Padmanabhan}}},
  \bibinfo{journal}{Phys. Rev. D} \textbf{\bibinfo{volume}{67}},
  \bibinfo{pages}{063504} (\bibinfo{year}{2003});
\bibinfo{author}{\bibfnamefont{T.}~\bibnamefont{{Padmanabhan}}},
  \bibinfo{journal}{Phys. Rev. D} \textbf{\bibinfo{volume}{66}},
  \bibinfo{pages}{021301} (\bibinfo{year}{2002}),
  \eprint[http://arXiv.org/abs]{hep-th/0204150}.

\bibitem[{\citenamefont{Medved}(2002)}]{medved02}
\bibinfo{author}{\bibfnamefont{A.~J.~M.} \bibnamefont{Medved}},
  \bibinfo{journal}{Phys. Rev.} \textbf{\bibinfo{volume}{D66}},
  \bibinfo{pages}{124009} (\bibinfo{year}{2002}), \eprint{hep-th/0207247}.


\bibitem[{\citenamefont{Birrell and Davies}(1982)}]{bd82}
\bibinfo{author}{\bibfnamefont{N.~D.} \bibnamefont{Birrell}} \bibnamefont{and}
  \bibinfo{author}{\bibfnamefont{P.~C.~W.} \bibnamefont{Davies}},
  \emph{\bibinfo{title}{Quantum Fields in Curved Space}}
  (\bibinfo{publisher}{Cambridge, UK: Cambridge University Press},
  \bibinfo{year}{1982}).

\bibitem[{\citenamefont{Sriramkumar and Padmanabhan}(2002)}]{sp02}
\bibinfo{author}{\bibfnamefont{L.}~\bibnamefont{Sriramkumar}} \bibnamefont{and}
  \bibinfo{author}{\bibfnamefont{T.}~\bibnamefont{Padmanabhan}},
  \bibinfo{journal}{Int. J. Mod. Phys.} \textbf{\bibinfo{volume}{D11}},
  \bibinfo{pages}{1} (\bibinfo{year}{2002}), \eprint{gr-qc/9903054}.

\bibitem[{\citenamefont{Boulware}(1975)}]{boulware75}
\bibinfo{author}{\bibfnamefont{D.~G.} \bibnamefont{Boulware}},
  \bibinfo{journal}{Phys. Rev.} \textbf{\bibinfo{volume}{D11}},
  \bibinfo{pages}{1404} (\bibinfo{year}{1975}).

\bibitem[{\citenamefont{Hartle and Hawking}(1976)}]{hh76}
\bibinfo{author}{\bibfnamefont{J.~B.} \bibnamefont{Hartle}} \bibnamefont{and}
  \bibinfo{author}{\bibfnamefont{S.~W.} \bibnamefont{Hawking}},
  \bibinfo{journal}{Phys. Rev.} \textbf{\bibinfo{volume}{D13}},
  \bibinfo{pages}{2188} (\bibinfo{year}{1976}).

\bibitem[{\citenamefont{Unruh}(1976)}]{unruh76}
\bibinfo{author}{\bibfnamefont{W.~G.} \bibnamefont{Unruh}},
  \bibinfo{journal}{Phys. Rev.} \textbf{\bibinfo{volume}{D14}},
  \bibinfo{pages}{870} (\bibinfo{year}{1976}).

\bibitem[{\citenamefont{Christensen and Fulling}(1977)}]{cf77}
\bibinfo{author}{\bibfnamefont{S.~M.} \bibnamefont{Christensen}}
  \bibnamefont{and} \bibinfo{author}{\bibfnamefont{S.~A.}
  \bibnamefont{Fulling}}, \bibinfo{journal}{Phys. Rev.}
  \textbf{\bibinfo{volume}{D15}}, \bibinfo{pages}{2088} (\bibinfo{year}{1977}).

\bibitem[{\citenamefont{{Wald}}(1984)}]{wald84}
\bibinfo{author}{\bibfnamefont{R.~M.} \bibnamefont{{Wald}}},
  \emph{\bibinfo{title}{{General relativity}}} (\bibinfo{publisher}{Chicago:
  University of Chicago Press}, \bibinfo{year}{1984}).

\bibitem[{\citenamefont{Rovelli}(1998)}]{rovelli98}
\bibinfo{author}{\bibfnamefont{C.}~\bibnamefont{Rovelli}},
  \bibinfo{journal}{Living Rev. Relativity} \textbf{\bibinfo{volume}{1}},
  \bibinfo{pages}{1} (\bibinfo{year}{1998}).

\bibitem[{\citenamefont{Padmanabhan}(2004)}]{pc-qnm}
\bibinfo{author}{\bibfnamefont{T.}~\bibnamefont{Padmanabhan}},
  \bibinfo{journal}{Class. Quant. Grav.} \textbf{\bibinfo{volume}{21}},
  \bibinfo{pages}{L1} (\bibinfo{year}{2004}), \eprint{gr-qc/0310027};
\bibinfo{author}{\bibfnamefont{T.~R.} \bibnamefont{Choudhury}}
  \bibnamefont{and}
  \bibinfo{author}{\bibfnamefont{T.}~\bibnamefont{Padmanabhan}},
  \bibinfo{journal}{Phys. Rev.} \textbf{\bibinfo{volume}{D69}},
  \bibinfo{pages}{064033} (\bibinfo{year}{2004}), \eprint{gr-qc/0311064}.

\end{thebibliography}

\end{document}